\documentclass[12pt]{article}
\usepackage{amsfonts}
\begin{document}
{\renewcommand{\thefootnote}{\fnsymbol{footnote}}

\begin{center}
{\LARGE  Symmetric States in Quantum Geometry}\\
\vspace{1.5em}
M.\ Bojowald\footnote{e-mail address:
{\tt bojowald@physik.rwth-aachen.de}} and H.~A.\ Kastrup\footnote{e-mail address: {\tt kastrup@physik.rwth-aachen.de}}\\
Institute for Theoretical Physics, RWTH Aachen\\
D--52056 Aachen, Germany\\
\vspace{1.5em}
\end{center}
}

\setcounter{footnote}{0}

\begin{abstract}
  Symmetric states are defined in the kinematical sector of
  loop quantum gravity and applied to spherical symmetry and
  homogeneity. Consequences for the physics of black holes and
  cosmology are discussed.
\end{abstract}

\section{Introduction}

In the kinematical sector of loop quantum gravity, states are
represented as functions on the space $\overline{\cal A}$ of
generalized connections on a principal fiber bundle $P(\Sigma,G)$.
Here, $G=SU(2)$ is the gauge group (which can be replaced with any
other compact Lie group in the following) and $\Sigma$ the space
manifold. All these states can be decomposed in terms of the spin
network basis which are special states associated with graphs in
$\Sigma$. Given a symmetry group $S$ acting on $\Sigma$, we can ask
for states which are symmetric with respect to that action and
therefore can be used to study the full theory in a simpler regime.
However, as the decomposition into spin networks shows, no ordinary
non-trivial state can be exactly symmetric: the discrete structure of
space breaks any continuous symmetry. Nevertheless, we can look for
symmetric {\em generalized\/} states which is automatic in our
definition \cite{SymmRed}:

\medskip

\noindent{\bf Definition.} {\em A symmetric state is a distribution on
  $\overline{\cal A}$ whose support contains only connections being
  invariant under the action of the symmetry group.}

\medskip

In order to describe symmetric states more explicitly, we need more
information about invariant connections. A well-known example is that
of an $SU(2)$-connection which is invariant under the rotation group:
\begin{eqnarray}\label{ansatz}
    A_a^i(x,\vartheta,\varphi)\,{\mathrm
    d}x^a\tau_i & = & a(x)\,{\mathrm d}x\,\tau_3+
    \cos\vartheta\,{\mathrm d}\varphi\,\tau_3
    +(-\phi_1(x)\,{\mathrm
    d}\vartheta+ \phi_2(x)\,\sin\vartheta\,{\mathrm
    d}\varphi)\tau_1\nonumber\\ 
    & & - (\phi_2(x)\,{\mathrm d}\vartheta+
    \phi_1(x)\,\sin\vartheta\,{\mathrm d}\varphi)\tau_2\,.
\end{eqnarray}
It can be decomposed into a reduced connection given by $a$ and scalar
fields $\phi_{1/2}$ which are functions on the radial reduced manifold
$B:=\Sigma/S$. This is in fact the general
situation \cite{Harnad,Brodbeck}: invariant connections can be
classified by a reduced connections and scalar fields on a reduced
bundle $Q$ over the reduced manifold $B:=\Sigma/S$. However, in
general the structure group of $Q$ is only a subgroup of $G$ (as in
the above example because the ansatz (\ref{ansatz}) is not gauge
invariant) which leads to a partial gauge fixing in the classical
description. We do not need to care about this partial gauge fixing
here because it can be undone in the quantum theory \cite{SymmRed}
where generalized connections and scalar fields are used which do not
rely on using a fixed fiber bundle.

So we arrive at a convenient representation of symmetric states as
functions on the space of generalized connections and scalar fields on
the reduced manifold. A basis for these states is given by spin
network states with Higgs field vertices in the reduced manifold with
gauge group $G$. The interpretation of these states is twofold: First,
we can use them in order to restrict the full theory to reduced models
given by a symmetry condition. This results in a loop quantization of
a mini- or midi-superspace model but has the advantage over usual
mini-/midi-superspace quantizations that loop quantization methods
can be directly applied resulting in a better comparison between a
model and the full theory. Second, due to their very definition
symmetric states can be seen as distributional states in
the full theory. Although they are in this sense idealized, they can
be approximated (in the weak topology) by ordinary states which have
the interpretation of {\em weave\/} states describing small
perturbations (which are necessary due to quantum fluctuations) around
a symmetric (e.g.\ homogeneous) geometry.

\section{Spherical Symmetry}

\begin{figure}[ht]
\begin{center}{\setlength{\unitlength}{1.5cm}
\begin{picture}(7,0.2)
    \put(0,0){\line(1,0){7}}
    \put(1,0){\circle*{0.1}}
    \put(2.5,0){\circle*{0.1}}
    \put(3.3,0){\circle*{0.1}}
    \put(5.1,0){\circle*{0.1}}
    \put(6,0){\circle*{0.1}}
    \put(4,-0.05){\line(0,1){0.1}}
    \put(4,0.2){\makebox(0,0){$x$}}
    \put(3.75,-0.18){\makebox(0,0){$e_-$}}
    \put(4.35,-0.18){\makebox(0,0){$e_+$}}
\end{picture}
}\end{center}
\caption{A spherically symmetric spin network.}
\label{fig:SphSymmSpin}
\end{figure}
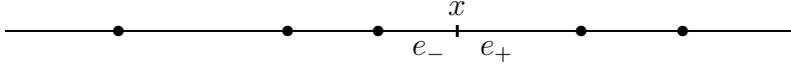

Spherically symmetric states are described by spin networks with Higgs
field vertices in a one-dimensional (radial) manifold $B$ (see Fig.\ 
\ref{fig:SphSymmSpin}). The Higgs field vertices can be interpreted as
representing edges being transversal to orbits of the symmetry group.
In the radial manifold a spherical surface is represented by a single
point in $B$, the radius $x$, and so it intersects a given spin network
state in only one point.

Denoting the radial dreibein components as $E_i$, the area of a
surface with radius $x$ is classically given by
\[
  A(x)=4\pi\sqrt{\delta^{ij}E_i(x)E_j(x)}\,.
\]
It can be quantized along the lines of the full theory \cite{Area} by
turning the dreibein components into functional derivatives with
respect to the radial connection components $A^i(x)$. Acting on a spin
network function $T$ only the edges $e_+$ and $e_-$ (which are assumed
to be outgoing in $x$) contribute via the holonomies
$h_{e_{\pm}}={\cal P} \exp\int_{e_{\pm}} {\rm{d}}x A^i\tau_i$. Using
the chain rule we obtain ($\gamma$ is the Immirzi parameter and
$l_{\rm{P}}$ the Planck length)
\begin{eqnarray*}
 \hat{E}_i(x)T & = & \frac{\gamma l_{\rm{P}}^2}{4\pi i}
  \sum_{\epsilon\in\{+,-\}}\int_{e_{\epsilon}}
  {\rm{d}} y\,\delta(x,y)(\tau_i\,h_{e_{\epsilon}})^A_B
  \frac{\partial}{\partial (h_{e_\epsilon})^A_B}\,T\\
 & = & \frac{\gamma l_{\rm{P}}^2}{4\pi}\sum_{\epsilon\in\{+,-\}}
  \frac{\epsilon}{2} J_{e_{\epsilon}}^iT
\end{eqnarray*} 
where $J_e=-iX_e$ is given by the right invariant vector field on the
copy of $SU(2)$ associated with the edge $e$. 

\begin{figure}[ht]
\begin{center}
{\setlength{\unitlength}{2.5cm}
\begin{picture}(4,5.2)
 \put(0,0){\line(1,0){4}}
 \put(0,0){\vector(0,1){5.2}}
 \put(0,1){\line(1,0){0.05}}
 \put(0,2){\line(1,0){0.05}}
 \put(0,3){\line(1,0){0.05}}
 \put(0,4){\line(1,0){0.05}}
 \put(0,5){\line(1,0){0.05}}
 \put(0.5,0.866){\line(1,0){1}}
 \put(0.5,1.414){\line(1,0){1}}
 \put(0.5,1.936){\line(1,0){1}}
 \put(0.5,2.449){\line(1,0){1}}
 \put(0.5,2.958){\line(1,0){1}}
 \put(0.5,3.464){\line(1,0){1}}
 \put(0.5,3.969){\line(1,0){1}}
 \put(0.5,4.472){\line(1,0){1}}
 \put(0.5,4.975){\line(1,0){1}}
 \put(-0.25,5){\makebox(0,0){$\displaystyle\frac{A}{\gamma l_{\rm{P}}^2}$}}
 \multiput(1.575,0.866)(0.1,0){9}{\line(1,0){0.04}}
 \multiput(1.575,1.414)(0.1,0){9}{\line(1,0){0.04}}
 \multiput(1.575,1.936)(0.1,0){9}{\line(1,0){0.04}}
 \multiput(1.575,2.449)(0.1,0){9}{\line(1,0){0.04}}
 \multiput(1.575,2.958)(0.1,0){9}{\line(1,0){0.04}}
 \multiput(1.575,3.464)(0.1,0){9}{\line(1,0){0.04}}
 \multiput(1.575,3.969)(0.1,0){9}{\line(1,0){0.04}}
 \multiput(1.575,4.472)(0.1,0){9}{\line(1,0){0.04}}
 \multiput(1.575,4.975)(0.1,0){9}{\line(1,0){0.04}}
 \put(2.5,0.866){\line(1,0){1}}
 \put(2.5,1.414){\line(1,0){1}}
 \put(2.5,1.732){\line(1,0){1}}
 \put(2.5,1.936){\line(1,0){1}}
 \put(2.5,2.280){\line(1,0){1}}
 \put(2.5,2.449){\line(1,0){1}}
 \put(2.5,2.598){\line(1,0){1}}
 \put(2.5,2.803){\line(1,0){1}}
 \put(2.5,2.828){\line(1,0){1}}
 \put(2.5,2.958){\line(1,0){1}}
 \put(2.5,3.146){\line(1,0){1}}
 \put(2.5,3.316){\line(1,0){1}}
 \put(2.5,3.351){\line(1,0){1}}
 \put(2.5,3.464){\line(1,0){1}}
 \put(2.5,3.669){\line(1,0){1}}
 \put(2.5,3.694){\line(1,0){1}}
 \put(2.5,3.824){\line(1,0){1}}
 \put(2.5,3.864){\line(1,0){1}}
 \put(2.5,3.873){\line(1,0){1}}
 \put(2.5,3.969){\line(1,0){1}}
 \put(2.5,4.012){\line(1,0){1}}
 \put(2.5,4.182){\line(1,0){1}}
 \put(2.5,4.217){\line(1,0){1}}
 \put(2.5,4.243){\line(1,0){1}}
 \put(2.5,4.330){\line(1,0){1}}
 \put(2.5,4.372){\line(1,0){1}}
 \put(2.5,4.386){\line(1,0){1}}
 \put(2.5,4.472){\line(1,0){1}}
 \put(2.5,4.535){\line(1,0){1}}
 \put(2.5,4.560){\line(1,0){1}}
 \put(2.5,4.690){\line(1,0){1}}
 \put(2.5,4.730){\line(1,0){1}}
 \put(2.5,4.739){\line(1,0){1}}
 \put(2.5,4.765){\line(1,0){1}}
 \put(2.5,4.835){\line(1,0){1}}
 \put(2.5,4.878){\line(1,0){1}}
 \put(2.5,4.894){\line(1,0){1}}
 \put(2.5,4.899){\line(1,0){1}}
 \put(2.5,4.975){\line(1,0){1}}
\end{picture}
}\end{center}
\caption{Level Splitting: $\sqrt{j(j+1)}$ (left) vs.\
  $\sum_p\sqrt{j_p(j_p+1)}$ (right).}
\label{fig:LevSplitt}
\end{figure}
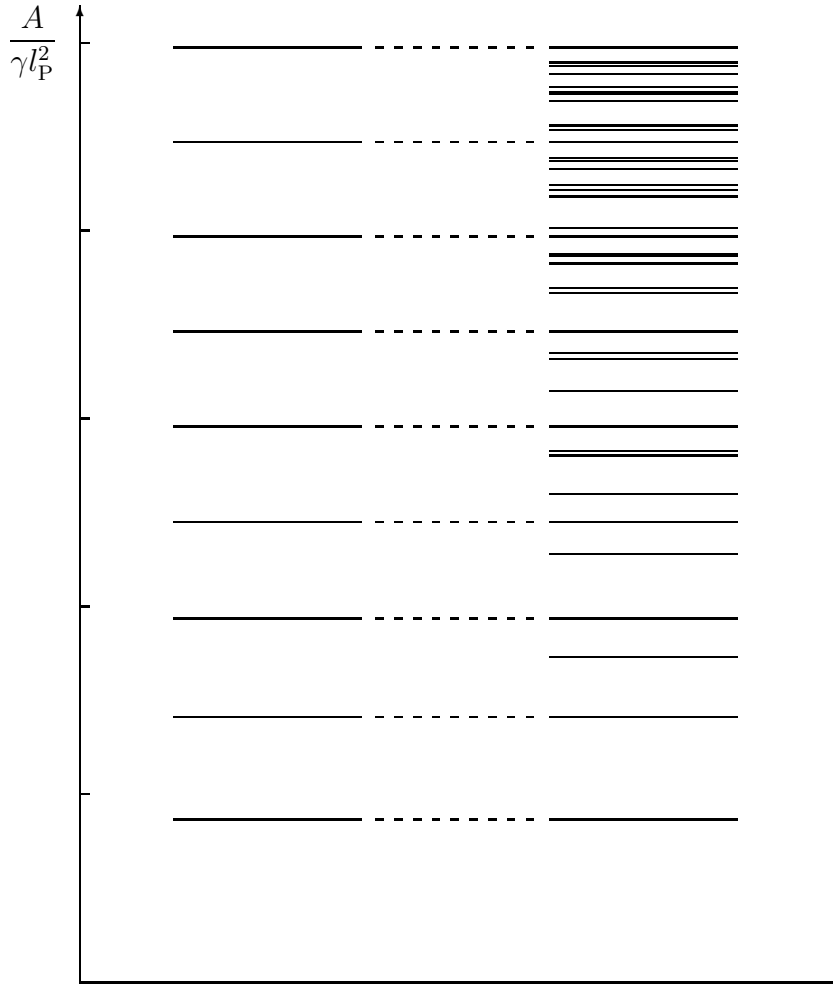

Inserted into the area $A(x)$ this yields
\[
 \hat{A}(x)={\textstyle\frac{1}{2}}\gamma\, l_{\rm{P}}^2
 \sqrt{(J_{e_+}-J_{e_-})^2}= {\textstyle\frac{1}{2}}\gamma
  l_{\rm{P}}^2\sqrt{2J_{e_+}^2+2J_{e_-}^2-(J_{e_+}+J_{e_-})^2}
\]
which can be simplified if $x$ is not a Higgs vertex because then
gauge invariance implies $J_{e_+}+J_{e_-}=0$. As compared to the area
operator in the full theory, the only difference is that the sum over
punctures disappears because in the spherically symmetric sector there
is only one puncture. Thus, we have the area spectrum (ignoring Higgs
vertices)
\begin{equation}
 A_j=\gamma l_{\rm{P}}^2\sqrt{j(j+1)}\quad,\quad j\in
 {\textstyle\frac{1}{2}} N_0\,.
\end{equation}

It is immediate to see that for large values of $j$ it is {\em
  equidistant\/} and compatible with the Bekenstein
spectrum \cite{Bekenstein} for the horizon area of spherically
symmetric black holes. On the contrary, the full area spectrum in loop
quantum gravity has an exponentially decreasing level
distance \cite{Area}. We can interpret the large difference of the two
spectra in the spherically symmetric and the non-symmetric regime as a
{\em level splitting\/} familiar from the spectroscopy of atoms:
breaking the spherical symmetry leads to a splitting of the levels
resulting in an almost dense spectrum (see Fig.\ \ref{fig:LevSplitt}).
A necessary requirement for this to happen is a huge degeneracy of the
levels in the spherically symmetric sector which is also expected from
thermodynamical considerations: in order to lead to a black hole
entropy proportional to the horizon area, the degeneracy of levels in
the Bekenstein spectrum has to grow exponentially \cite{BekMuk}.
However, in quantum geometry spherically symmetric states are
distributional and so their degeneracy is not well defined which
prohibits a simple counting of states.

\section{Loop Quantum Cosmology}

In specializing the framework to homogeneous or even isotropic states
we have the basics of loop quantum cosmology \cite{cosmoI}. In this
case, the reduced manifold is a single point and so homogeneous states
are defined only in terms of scalar fields (point
holonomies \cite{FermionHiggs}). For Bianchi models (anisotropic) we
have three independent point holonomies which can be visualized as
being associated with three closed edges meeting in a single
$6$-vertex. For isotropic models there is only one point holonomy, but
here the concept of spin networks has to be generalized in order to
describe all states: the configuration space is no longer a group due
to a condition on the scalar fields in an invariant connection if the
symmetry group has a non-trivial isotropy subgroup. Therefore, the
Peter--Weyl theorem which tells us that functions on a compact group
can be expanded into functions associated with matrix elements of all
its irreducible representations (which are just spin networks if the
group is a product of some copies of $SU(2)$) does not apply and a
substitute for the expansion has to be found by different means. In
the simple case of gauge invariant isotropic states, which depend on
only one parameter, it turns out that one can describe all those
states as spin networks associated with a single closed edge but
possibly with an insertion in the vertex \cite{cosmoII}. In this way,
the number of states is doubled as compared to a naive expectation.

\subsection{Volume}

Again, we can observe the phenomenon of {\em level splitting}, this
time for the volume operator \cite{cosmoII}. For Bianchi models, the
situation is similar to the area operator in the spherically symmetric
sector: the sum over vertices disappears because there is only one
$6$-vertex to which the point holonomies are attached. For isotropic
states, the operator simplifies so much that the complete volume
spectrum can be computed explicitly ($V_0$ is an arbitrary constant
entering via a homogeneous auxiliary metric):
\begin{equation}
 V_j=\gamma^{\frac{3}{2}}\, V_0^{-\frac{1}{2}}\, l_{\rm{P}}^3\,\sqrt{j
   (j+{\textstyle\frac{1}{2}}) (j+1)}
 \quad,\quad j\in{\textstyle\frac{1}{2}}N_0\,.
\end{equation}

The fact that homogeneous states are distributional also implies that
there is a discrepancy between mini-superspace quantizations and
approximations (by weaves) in the full theory: even small
inhomogeneous perturbations cause a transition to the full volume
spectrum. This is in contrast to the treatment of inhomogeneities in
more standard quantum cosmological models \cite{Halliwell} where
symmetric and slightly perturbed geometries are smoothly connected.

\subsection{Dynamics}

In cosmological models there is a familiar procedure to study
intrinsic dynamics by introducing the volume as internal time. The
volume quantization in quantum geometry suggests that such a procedure
in loop quantum cosmology leads to a discrete time. This can in fact
be made more precise by implementing the intrinsic dynamics into a
loop quantization of cosmological models. To that end we need, in the
first place, a quantization of the Hamiltonian constraint for
cosmological models \cite{cosmoIII} which can be derived using the key
steps of the quantization in the full theory \cite{QSDI} with suitable
adaptations in order to respect the symmetry.

The result for Bianchi models is a constraint operator which looks
very similar to that in the full theory: the Euclidean part is
\begin{equation}\label{HamEuclBianchiQuant}
 \hat{{\cal H}}^{({\rm{E}})}[N]=-4i(\gamma l_{\rm{P}}^2)^{-1} V_0 N
   \sum_{IJK}\epsilon^{IJK} 
 {\rm{tr}}\left( h_Ih_Jh_I^{-1}h_J^{-1} h_{[I,J]}^{-1} [h_K,\hat{V}]\right)
\end{equation}
using the volume operator $\hat{V}$ of Bianchi models mentioned above
and $h_I$, $1\leq I\leq 3$ are the three point holonomies interpreted
as multiplication operators in a connection interpretation. We also
defined
\[
 h_{[I,J]}:=\prod_{K=1}^3 (h_K)^{c^K_{IJ}}
\]
which is the only place where the constraint operator depends on the
Bianchi type (via the structure constants $c^K_{IJ}$). Once we have
the Euclidean constraint, the Lorentzian one can be derived completely
analogously to the full theory:
\begin{equation}\label{HamBianchiQuant}
 \hat{{\cal H}}[N]=8i(1+\gamma^2)(\gamma l_{\rm{P}}^2)^{-3}V_0 N
 \epsilon^{IJK} {\rm{tr}} \left([h_I,\hat{K}] [h_J,\hat{K}]
   [h_K,\hat{V}]\right)- \hat{{\cal H}}^{(\rm{E})}[N]
\end{equation}
using the extrinsic curvature
\[
  \hat{K}=i\hbar^{-1}\left[\hat{V},\hat{{\cal H}}^{(\rm{E})}[1]\right]\,.
\]

In order to find a dynamical interpretation of the constraint equation
we have to transform from the connection to a dreibein representation
and to select an internal time. The transformation to a dreibein
representation can be done by expanding an arbitrary homogeneous state
$c$ in terms of spin network states labeled by elements $L$ of an
index set ${\cal I}$ (e.g., three edge spins and the $6$-vertex
contractor for Bianchi models)
\[
  c(A)=\sum_{L\in{\cal I}}c_LT_L(A)
\]
and interpreting the coefficients $c_L$ as components of the state in
the dreibein representation.

Unfortunately, the volume operator for Bianchi models is quite
complicated and cannot be diagonalized explicitly. Therefore, it is
awkward to introduce the volume as internal time. However, we can also
choose simply one of the edge spins $n$ as internal time (being
related to a diagonal component of the homogeneous metric) and still
have a discrete time label. Because the Hamiltonian constraint
(\ref{HamBianchiQuant}) contains holonomies as multiplication
operators there are always terms with different `time' labels $n$ in
the action of $\hat{\cal H}$ on a state. In the dreibein
representation we can write
\begin{equation}
 (\hat{{\cal H}}c)_n= \sum_{i=-\frac{\omega}{2}}^{\frac{\omega}{2}}
    (H_ic)_{n+\frac{i}{2}}
\end{equation}
by collecting on the right hand side all contributions with a fixed
spin $n$ (the $H_i$ are operators acting on the remaining labels which
are suppressed in our notation). The order $\omega$ is given by twice the
maximal number of holonomies acting on the selected edge labeled by
$n$ which appear in a term of the Hamiltonian constraint operator.

We thus arrive at a Wheeler--DeWitt equation of the form
\[
 \sum_{i=-\frac{\omega}{2}}^{\frac{\omega}{2}}(H_ic)_{n+\frac{i}{2}}=0\quad
 \mbox{ for all }\quad n\in{\textstyle\frac{1}{2}} N_0
\]
which is a {\em discrete} evolution equation \cite{cosmoIV} rather than
a differential equation as in standard quantum cosmology. Given a set
of initial conditions for small $n$ (the number given by the order
$\omega$) one can compute the evolution as long as the highest
operator $H_n$ is invertible.

In these models, one can also show that the {\em physical} volume
spectrum defined using evolving observables is identical to the
kinematical one; in particular, it is discrete.

\section*{Acknowledgments}

M.\ B.\ is grateful to the DFG-Graduierten-Kolleg `Starke und
elektroschwache Wechselwirkung bei hohen Energien' for a PhD
fellowship and travel grants.

\end{document}